\documentclass{rspublic}

\usepackage{graphicx}
\usepackage{subfigure}

\begin{document}

\title[Scaling in Binary Fluid Lattice Boltzmann]{Physical and Computational Scaling Issues in Lattice Boltzmann Simulations of Binary Fluid Mixtures}

\author[M. E. Cates and others] 
{M. E.  Cates$^1$, J.-C. Desplat $^{2}$, P. Stansell$^1$, A. J. Wagner$^{3}$,\\
K. Stratford$^2$, R. Adhikari$^1$, I. Pagonabarraga$^4$}

\affiliation{ 
  $^1$ School of Physics, University of Edinburgh, JCMB King's Buildings,\\ Edinburgh EH9 3JZ, UK\\
  $^2$ EPCC, University of Edinburgh, King's Buildings, Edinburgh EH9 3JZ, UK\\
  $^3$ Department of Physics, North Dakota State University, Fargo, ND 58105, USA \\
  $^4$ Departament de Fisica Fonamental, Universitat de Barcelona,\\ Av. Diagonal 647, 08028 Barcelona, Spain}

\label{firstpage}

\newcommand\etal{\textit{et al. }}
\renewcommand\Re{\mbox{\it Re}}
\newcommand\Pe{\mbox{\it Pe}}
\newcommand\Ca{\mbox{\it Ca}}
\newcommand\Bo{\mbox{\it Bo}}
\newcommand\out[1]{}

\maketitle

\begin{abstract}{Lattice Boltzmann, scaling, boundary conditions, shear flow, complex fluids, colloids, demixing}
We describe some scaling issues that arise when using lattice Boltzmann methods to simulate binary fluid mixtures -- both in the presence and in the absence of colloidal particles. Two types of scaling problem arise: physical and computational. Physical scaling concerns how to relate simulation parameters to those of the real world. To do this effectively requires careful physics, because (in common with other methods) lattice Boltzmann cannot fully resolve the hierarchy of length, energy and time scales that arise in typical flows of complex fluids. Care is needed in deciding what physics to resolve and what to leave unresolved, particularly when colloidal particles are present in one or both of two fluid phases. This influences steering of simulation parameters such as fluid viscosity and interfacial tension. When the physics is anisotropic (for example, in systems under shear) careful adaptation of the geometry of the simulation box may be needed; an example of this, relating to our study of the effect of colloidal particles on the Rayleigh-Plateau instability of a fluid cylinder, is described. The second and closely related set of scaling issues are computational in nature: how do you scale up simulations to very large lattice sizes? The problem is acute for systems undergoing shear flow. Here one requires a set of blockwise co-moving frames to the fluid, each connected to the next by a Lees-Edwards like boundary condition. These matching planes lead to small numerical errors whose cumulative
effects can become severe; strategies for minimising such effects are discussed. \end{abstract}

\section{Introduction}

The lattice Boltzmann equation (LB equation, or LBE) is a widely used lattice formulation of fluid mechanics (Succi, 2001). It offers a faithful discretisation of
the Navier Stokes equation of isothermal, near-incompressible fluid flow, and is very well adapted to parallel computation (Amati \etal 1997). The
LBE approach is particularly adapted to simulating mesoscopic problems (Coveney \& Succi, 2001). These include, for example, porous medium flows, and flows of
complex and multicomponent fluids with microstructure
(Manz \etal 1999; Swift \etal 1995; Kendon \etal 1999, 2001; Pagonabarraga \etal 2001; Nekovee \etal 2000; Ladd, 1994a, 1994b; Love \etal 2003).

For binary fluids, the LB equations read (with timestep $\Delta t = 1$)
\begin{equation}
f_i({\bf x} + {\bf c_i}, t+1) = \sum_jL_{ij}(f_j({\bf x},t) -f_j^0({\bf x},t))
\end{equation}
\begin{equation}
g_i({\bf x} + {\bf c_i}, t+1) = \sum_jM_{ij}(g_j({\bf x},t) -g_j^0({\bf x},t))
\end{equation}
Here ${\bf x}$ is a lattice position; ${\bf c}_i$ the local velocity set corresponding to a set of near-neighbour vectors of the lattice (selected to recover rotational and Galilean invariance properties); $f_i$ the local distribution function for molecular velocities at the given site; and $f_i^0(\rho,{\bf v})$ its equilibrium value (dependent on local fluid density and flow velocity). The distribution function $g_i$ controls the evolution of a binary fluid order parameter $\phi$; its equilibrium value $g_i^0$ encodes both a well-chosen free energy functional, $F(\phi,\nabla\phi) = A\phi^2/2 + B\phi^4/4 + \kappa(\nabla\phi)^2/2$, and an order parameter mobility (Swift \etal 1995; Kendon \etal 2001). The matrices
$L_{ij}$ and $M_{ij}$ are collision operators; $L_{ij}$ is often chosen as a single-relaxation time (lattice BGK) matrix $-\delta_{ij}/\tau$, in which case the fluid viscosity is $\eta = \rho c_s^2(\tau-1/2)$ with $c_s$ the sound speed (controlled by the choice of ${\bf c}_i$).  In our work $M_{ij}$ is chosen of BGK form with $\tau = 1$, so that $g_i$ is reset to $g_i^0$ every timestep.
The density fields $\rho = \sum_i f_i$ and $\phi = \sum_i g_i$ are the zeroth moments of the distribution functions; the fluid mass fluxes obeys $\rho{\bf v} = \sum_if_i{\bf c}_i$ and $\rho\phi{\bf v} = \sum_ig_i{\bf c}_i$. Finally, the second moment of $\sum_if_i{\bf c}_i{\bf c}_i$ determines the kinetic stress tensor (or momentum flux tensor).

These standard LB equations neglect thermal noise and lead to deterministic (zero temperature) evolution of the fluid velocity field ${\bf v}$ and order parameter $\phi$; the continuum level equations to which they correspond are the isothermal Navier Stokes equation for ${\bf v}$, properly coupled to an advection-diffusion equation for $\phi$. Lattice Boltzmann must only be used at low Mach number when the fluid density $\rho$ remains nearly constant. It is possible to add thermal noise to the fluid either at the level of low wavevector hydrodynamics via the stress tensor (Ladd, 1994a, 1994b), or consistently across all wavevectors by developing discrete Langevin equations of which the above are the deterministic limit (R. Adhikari, K. Straford, A. Wagner and M. E. Cates, to be published). When noise is referred to below, it has been included by the latter route, but in the fluid momentum sector only ($f_i$ but not $g_i$), corresponding to a thermal fluid in which the order parameter dynamics are `cold', so that thermal capillary waves at the fluid-fluid interface are neglected. Work on the full incorporation of Langevin noise into a binary fluid model is underway. Meanwhile, this method allows incorporation of Brownian colloidal particles within a binary fluid system.

\section{Physical scaling issues}
We first address issues of physical scaling, that is, how to choose parameter values in LB that map onto those of the real world. At one level this is trivial, at least in our own implementation of LB (Desplat \etal 2001) in which thermodynamic quantities such as interfacial tension $\sigma = (8\kappa|A^3|/9B^2)^{1/2}$ between two fluids follow from a well defined free energy functional. The thermodynamic and kinetic properties of an LB fluid (or fluid mixture) are then defined, through the choices made above for $L_{ij}, M_{ij}$ and $g_i^0$, in terms of lattice units for mass, length and time. The mass unit contains an arbitrary scale factor (we set $\rho = 1$) whereas lattice units for length and time are directly linked to the lattice constant and timestep in the simulation. In principle there is no problem, therefore, choosing all the thermodynamic and kinetic parameters in LB to correspond to those of a real physical system, by an appropriate mapping of lattice units onto real ones.

In practice, of course, this is rarely possible. Setting aside various stability issues (Kendon \etal 2001), the run times and system sizes required would be simply too big, as illustrated by the examples that follow. Accordingly, LB is generally interpreted as a `mesoscale' simulation method which throws away local information so as to achieve much larger length- and timescales than could be achieved with, say, molecular dynamics. On the other hand, LB does not throw away as much information as does computational fluid dynamics (CFD), which discretises directly the Navier Stokes equation and treats any interfaces as structureless.
 
\subsection{What is the difficulty?}

Consider the problem of disconnection, in which a droplet of one fluid surrounded by another (for simplicity, of equal viscosity) breaks into two droplets via a pinchoff event. (This was studied carefully in two dimensions by Wagner \etal 2003b). At the CFD level this can be modelled using a Navier-Stokes continuum, with a structureless interface between the fluids, characterised only by an interfacial tension $\sigma$. But this problem is singular (Eggers 1997) so that direct CFD methods fail, unless explicit measures are taken to identify incipient reconnection of the interface and deal with it `by hand'. Such failure of CFD is inevitable, because the ultimate pinchoff of a thin neck is controlled by molecular, not continuum, physics. On the other hand, LB maintains a coarse-grained representation of the molecular physics through its treatment of distribution functions on the lattice. This leads to the smooth pinchoff of fluid necks at the point where their diameter approaches the interfacial width (generally a couple of lattice spacings). When LB is used for binary fluid simulations, this removes the singularities that would obstruct a direct approach, and does so in a physically `realistic' way (effectively by molecular dissolution of the minority fluid under the influence of Laplace pressure). 

On the other hand, the LB treatment of this and similar problems is rarely {\em fully} realistic (Cates \etal 2004): the mechanism is valid, but the scale at which it sets in is not. To see this, note that in the real world the molecular cutoff point is not reached until the diameter of a fluid neck is at sub-nanometre level. In LB the molecular physics takes over at the scale of the lattice constant, and if this was always linked to a physical scale of (say) 0.1nm, there would be no possibility of simulating systems larger than 100nm across, even with the largest computers available. (These can run $1024^3$ lattices (Harting \etal 2004) although $128^3$ is routinely used in our own work.) Fortunately this limitation, which would restrict LB to broadly the same parameter domain as molecular dynamics, is more illusory than real. For, so long as the pinchoff process itself does not dominate the physics, it does not matter whether or not this is resolved in a {\em fully} realistic manner. 
The advantage of having a realistic mechanism remains: because the pinchoff process is physically admissible (albeit for some set of fluid parameters that may remain strictly hypothetical in the laboratory), it is less likely to lead to gross artefact than an arbitrary but well-intentioned numerical discretisation. Having said this, gross artefacts are sometimes easier to detect and control than subtle ones; LB is quite prone to the latter, and careful testing for systematic errors is always needed (see e.g. Kendon \etal 2001 for a fuller discussion).

\subsection{Binary fluid demixing}

A good example of the mesoscale use of LB simulation is in the coarsening of fluid domains after spinodal decomposition in a pair of fully symmetric binary fluids (Kendon \etal 2001). The domain length scale $L(t)$ (appropriately defined; see below) at time $t$ after quench in such a system is widely held to obey a scaling of the form (Bray 1994, 2000; Onuki 2002)
\begin{equation}
L(t)/L_0 = f(t/T_0)
\end{equation}  
where $L_0 = \eta^2/(\rho\sigma)$ is a characteristic length, $T_0 = \eta^3/(\rho\sigma^2)$ a characteristic time, and $\eta$ and $\rho$ the
(equal) viscosities and interfacial tension of the fluids. Since $L_0$ and $T_0$ involve only macroscopic quantities, the observation of such scaling in experiments (Onuki 2002) is evidence that the molecular physics of pinch-off does not matter. Moreover, such scaling implies that results for LB runs with widely different $\eta$ and $\sigma$ should collapse onto each other when scaled onto an appropriate plot. This is indeed seen (Pagonabarraga \etal 2002), showing that in LB, as in the real world, the details of pinchoff are irrelevant. However, careful measures must first be taken to eliminate artefacts such as, for example, excessive interdiffusion of the two fluids on the domain scale $L(t)$. These measures amount to demanding that a specific hierarchy of physical effects is obeyed: on the time scale of coarsening, diffusion must be easily accomplished across the width of a fluid-fluid interface (a few lattice sites), so as to maintain an equilibrium surface tension, but simultaneously diffusion by the same mechanism must be negligible at the domain scale. 

\subsection{Dimensionless control parameters}

In applying LB methods to a wider class of problems in complex fluids modelling, it is wise to be explicitly conscious of the hierarchy of scales that the real-world problem presents. Only by respecting this hierarchy within the simulation can one expect realistic results. In fact, reckoning with the hierarchy in physical terms is needed, even to decide whether the problem lies within the scope of LB for any given computational resource. The best way to develop these ideas is using the dimensionless numbers beloved of fluid-mechanicists, who have long since covered a similar territory in developing scaling relations between one set of experimental parameters and another. When done well, these analyses pay as much attention to discussing what can be neglected as to what must be included (Batchelor 1967, Faber 1995). 

In the binary fluid coarsening problem, the two relevant numbers are the capillary number $\Ca = \eta \dot L/\sigma = \dot L T_0/L_0$ and the Reynolds number $\Re = L \dot L T_0/L_0^2$. The first measures the ratio of viscous to interfacial forces; the second the ratio of inertial to viscous forces. (A third combination $\Re\,\Ca$ therefore measures the ratio of inertial to interfacial forces.) At early times $t \ll t^*$, viscous and interfacial forces are in balance so that $\Ca\sim 1$ (giving $L(t)/L_0 \sim t/T_0$) and $\Re\ll 1$. At late times $t \gg t^*$, inertial and interfacial forces balance so that $\Re\,\Ca \sim 1$ (giving $L/L_0 \sim (T/T_0)^{2/3}$) and $\Ca\ll 1$. Here $t^*$ is a crossover time `of order' $T_0$; our LB work fixes the prefactors in the two coarsening laws and shows that in fact $t^*\simeq 10^4T_0$ (Kendon \etal 1999), perhaps stretching the interpretation of the term `of order'.

Intriguingly, careful investigation of one relatively simple-looking extension to this problem --- binary fluid demixing under gravity in the presence of a continuous temperature ramp --- throws up at least three new dimensionless numbers (Cates \etal 2003) to add to the many that already arise in the scaling of complex fluids. Indeed, not just Reynolds, but Peclet, Schmidt, Rayleigh, Bond, Prandtl, Benard, Weissenberg, and more, can feature under different experimental conditions. (Such surnames give no direct clue as to the actual physics of these numbers; names like `inertial/viscous ratio' would be a lot more informative than `Reynolds number', but convention precludes them.) The relatively subtle hierarchies that arise in problems such as the temperature-ramped binary fluid pose a severe test on the LB methodology. In that particular case, LB work remains in progress, with preliminary results described by Wagner \etal (2003a).

\subsection{Colloids in binary fluids}
The addition of colloidal particles to a binary fluid system raises further scaling considerations. Our LB algorithm for simulating colloids is based on that of Nguyen \& Ladd (2002); it is outlined briefly by Cates \etal (2004) and full details will be published elsewhere (K. Stratford \etal in preparation). 

\begin{figure}
\begin{center}
\includegraphics[width=12cm]{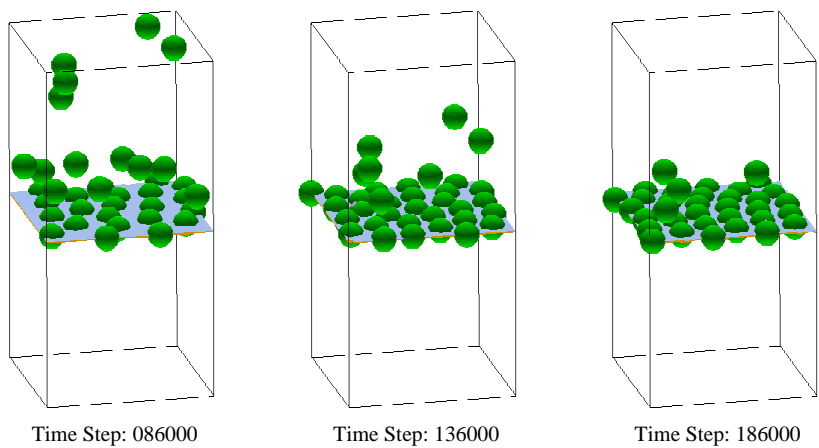}
\end{center}
\caption{\label{bench}Snapshot configurations for the benchmark sedimentation problem with $\Bo^{-1} = 280$ and $\Re = 0.02$. MOVIE IN ONLINE JOURNAL (5.5MB):
http://www.ph.ed.ac.uk/~mec/splash3.avi} 
\end{figure}

Consider first a simple geometry in which a suspension of colloids, at low volume fraction, sediments under gravity within a stratified pair of layers of two immiscible fluids of equal viscosity. Figure \ref{bench} shows this process. The colloidal particles are initially placed at random; they then fall under gravity and, for the parameters selected here, end up attached to the interface. The interfacial tension between fluids is $\sigma$; the solid-fluid tensions are equal, so that particles at the interface between fluids have a 90 degree wetting angle; both fluids have viscosity $\eta$; the colloid radius is $a$.
The relevant dimensionless numbers are those of Reynolds, Peclet and Bond:
\begin{equation}
\Re = \frac{v_{sed} a \rho}{\eta}
\end{equation}
\begin{equation}
\Pe = \frac{v_{sed} a}{D}
\end{equation}
\begin{equation}
\Bo = \frac{9}{2}\frac{v_{sed}}{v_{0}} = \frac{\Delta\rho g a^2}{\sigma}
\end{equation}
Here $v_{sed} = \Delta mg/6\pi\eta a$ (with $\Delta m = 4\pi a^2\Delta\rho/3$) is the sedimentation speed of a colloid, and
$D = k_BT/6\pi\eta a$ its diffusion constant; $v_0 = \sigma/\eta$ is a characteristic velocity associated with coarsening (it obeys $v_0=L_0/T_0$ as defined above). The sedimentation velocity is fixed by balancing gravity against
viscosity; accordingly the Bond number (ratio of gravity to interfacial force) is essentially the same as the capillary number, which we defined above as the ratio of viscous to interfacial force. [Note that Cates \textit{et al.} (2004) used an inverted definition of capillary number, so that $\Ca$ in that paper means $9\Bo^{-1}/2$; to avoid further confusion, we use the Bond number, rather than capillary, in what follows.] 

For typical colloids in binary fluid mixtures one has
$\Re \simeq 10^{-7} (a/a_\mu)^3$;  $\Pe\simeq (a/a_\mu)^4$; and $\Bo \simeq 10^{-8}(a/a_\mu)^{2}$, where $a_\mu \equiv 1\mu$m. Thus a colloid of, say, $a = 10\mu$m has negligibly small inertial effects (small $\Re$); negligibly small diffusion (large $\Pe$); and a very large capillary force attaching it to the fluid-fluid interface, easily capable of holding it there against gravity (very small $\Bo$). For $a \le 100$nm, $\Re$ and $\Bo$ are still negligible, although $\Pe$ is now also small: diffusion is important. But even for 5nm colloids, the combination
$\Pe/\Bo$ remains large; this represents the ratio of interfacial to thermal forces. Thus, for all reasonable interfacial tensions and colloid sizes, neutrally wetting colloidal particles adsorb quasi-irreversibly to the fluid-fluid interface, and {\em diffusive} detachment does not take place (Aveyard \etal 2003, Binks 2002).

Suppose, therefore, that we wish to simulate $a = 0.1\mu$m radius neutrally-wetting colloids in an oil water mixture. Thermal noise can be neglected; and as outlined above in the case of gravity (but true more generally) for typical colloidal problems $\Re$ is extremely small. However, LB works by solving dynamically a discretisation of the full hydrodynamic equations. Thus, as explained in more detail by Cates \etal (2004), small Reynolds number can only be achieved by having a colloidal velocity that is, at most, of order $\Re$ in lattice units. For a run such as that of Figure \ref{bench} in which colloids move a distance of order the simulation box (say 100 lattice sites), the number of timesteps required for $a/a_\mu=0.1$ then obeys $N\sim 100/\Re \simeq 10^{12}$. This run-time is roughly a million times longer than anything practically achievable on the world's fastest computers. 

Fortunately, as previously explained, `fully' realistic simulations (in which lattice parameter values map directly onto those of the real world) are not the goal of mesoscale LB. Indeed, for the present problem, fluid mechanical lore (Batchelor, 1967) asserts that for an isolated sphere, all Reynolds numbers less than unity are practically equivalent. This is not quite enough to show that LB is accurate, since the time discretisation error in LB enters formally at the same order as the inertial terms that $\Re$ describes, but with a different prefector. However, Cates \etal (2004) confirm that any $\Re \le 0.02$ is clearly negligible and $\Re \le 0.1$ probably adequate, bearing in mind typical systematic errors of a few percent (Kendon \etal 2001) from each of several other sources. In summary, however small $\Re$ is in a specified real-world problem, we should not waste time making it smaller than we need to, to gain accurate simulation results for that problem.

By the same token, for many mesoscale simulation purposes one expects that any Bond number less than some small value is equivalent to any other. Of course, this is not true if one looks at the detailed deformation of the fluid interface close to a particle suspended there against gravity, but for multi-colloid simulations that is unlikely to be well resolved in LB anyway. Indeed, in such simulations it is normal to use colloidal radii of about three lattice units or even less (Ladd 1994a, 1994b; Nguyen \& Ladd, 2002), whereas the full width of the fluid-fluid interface is usually one or two lattice units (Kendon \etal 2001). Although the interfacial tension in LB cannot be raised to very large values without risk of instability, it is not difficult to achieve Bond numbers of order $3\times 10^{-4}$ (as used in Figure \ref{bench}). 

\begin{figure}
\begin{center}
\includegraphics[width=10cm]{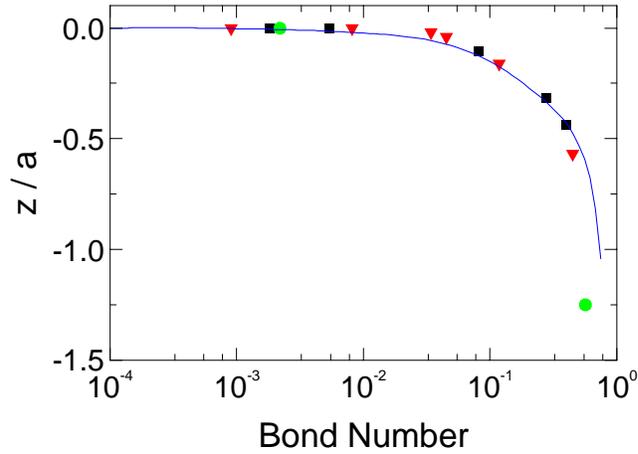}
\end{center}
\caption{\label{bond} Displacement $z/a$ under gravity of isolated particles of radius $a$ suspended at an interface. System size $48\times 48\times 96$, particle radii in lattice units 2.3 (circles), 3.71 (triangles), 4.77 (squares).
The solid line is Derjaguin's equation (strictly valid at small Bond number only). The LB simulation parameters vary between runs but all are among those benchmarked by Kendon \etal (2001; their Table 3).} 
\end{figure}

\begin{figure}
\begin{center}
\includegraphics[width=12cm]{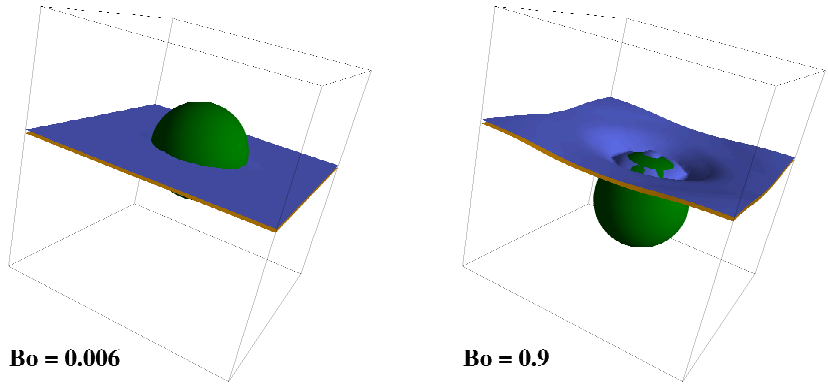}
\end{center}
\caption{\label{bond2} Interfacial configuration for a colloidal particle at low and high Bond number. The latter is en-route to detachment. Particle size $a = 4.77$. Displacement $z/a$ under gravity of particles of radius $a$ suspended at an interface. System size $48\times 48\times 96$, particle radii in lattice units 2.3 (circles), 3.71 (triangles), 4.77 (squares).} 
\end{figure}

Figure \ref{bond} shows the dependence on $\Bo$ of the mean particle displacement from the zero-gravity equilibrium position, for particles suspended from an interface (as in Figure \ref{bench}, but now with a single particle only) on varying the strength of gravity and/or interfacial tension. This is compared with a theoretical prediction based on the result of Derjaguin (1946) that should be accurate at low $\Bo$. Agreement with the latter is excellent; moreover, bearing in mind other sources of systematic error, so long as $\Bo \le 10^{-2}$ or certainly $\Bo \le 10^{-3}$, such displacements are close enough to zero for these $\Bo$ values to serve as an adequate numerical substitute for fully realistic values of order $10^{-8}$. (Interestingly, one such source of error involves spurious momentum currents in the plane of the interface; these are small --- $\le 10^{-5}$ in lattice units --- but mean that particles do not come to a complete halt on the interface even, as here, in the absence of thermal noise.) Figure \ref{bond2} shows a particle at the interface for $\Bo = 0.006$ and $\Bo = 0.9$. In the first case the interface is effectively flat; in the second, the particle is en-route to detachment. Our preliminary simulations find a critical Bond number for detachment between 0.6 and 0.9, with a small dependence on particle radius $a$; the theoretical value is $3/4$. Much higher accuracy would require larger particles, and certainly one would expect to use these in any problem where the local physics of detachment was dominant.
 
\section{Computational scaling issues: Sheared binary fluids}

Computational scaling issues are in one sense purely algorithmic: the concern is how to scale up the simulation to deal with larger systems sizes and thus reduce discretisation errors and finite-size effects. We deal with these questions here, using as an example the important problem of symmetric binary fluids under shear. This problem has been widely studied by simulation (Cates \etal 1999, Wagner \& Yeomans 1999, Harting \etal 2004, Corberi \etal 1998, Lamura \etal 2002) and also by theory (Bray 2003, Cates \etal 1999, Doi \& Ohta, 1991). But so far definitive LB results have proved elusive, in large part because these require very large system sizes before finite size effects can be brought under control.

The scientific issue is relatively simple: does steady shear prevent the demixing of binary fluids, and if so, how do the resulting steady-state length scales for the fluid domains depend on shear rate? `Length scales' is plural, because theory shows that in principle the characteristic domain sizes measured in the three directions of velocity ($x$), velocity gradient ($y$), and vorticity ($z$) can differ by arbitrarily large factors. Indeed, in one type of approximate theory, in which the shear flow is assumed perfectly laminar at all times (Rapapa \& Bray 1999, Bray 2003) all three of these length scales diverge in time, but with different power laws. Simple scaling arguments, which ignore anisotropy but allow for a fluctuating velocity profile, do not rule out a divergence with time of the (now single) domain scale $L(t)$ under shear. However, they also allow at least two possible scenarios for its saturation. 
The argument of Doi \& Ohta (1991) proposes that saturation occurs when the capillary number is of order unity; this gives $\Ca \sim \dot \gamma L\eta/\sigma \sim 1$ so that the domain scale in steady state obeys $L\sim\sigma/(\eta\dot\gamma)$. In contrast, Cates \etal (1999) argue that coarsening might cease only when the Reynolds number is of order unity; this gives $\Re \sim \rho \dot\gamma L^2/\eta \sim 1$ or $L\sim\eta^{1/2}(\rho\dot\gamma)^{-1/2}$. 

There are two separate reasons why these arguments still have not been definitively tested by LB (in either two or three dimensions). The first is purely to do with numerical scale-up and involves the need to subdivide the LB lattice to achieve large shear rates. The second concerns finite size effects: specifically, what shape of simulation box should be used, and how can one know this {\em in advance} of doing the simulation? We address these issues in turn.

\subsection{Multiple Lees Edwards planes}
Until recently the best way to apply steady shear to a binary LB fluid was to create a simulation box with solid walls on two parallel faces (and periodic boundary conditions on the rest). At these solid walls, a specific velocity can be imposed. Unfortunately this method of applying shear poses a severe obstacle to computational scale-up, as follows. In LB, the local fluid velocity relative to the lattice cannot be allowed to get too large: $v_{max} = c_s/10\simeq 0.06$ (in lattice units) is a good rule of thumb. (Beyond this, instability and/or severe breakdown of Galilean invariance sets in.) Applying this maximum velocity with opposite signs at the two walls gives a maximum shear rate $\dot \gamma = 2v_{max}/\Lambda_y$, with $\Lambda_y$ the separation of the walls. This makes it impossible to increase the system size at fixed shear rate; indeed the achievable $\Re$ increases only linearly with $\Lambda$, while $\Ca$ does not increase at all!

A breakthrough for this problem was made by Wagner \& Pagonabarraga (2002) who developed Lees-Edwards boundary conditions (LEBC) for LB. In molecular dynamics, LEBC (Lees \& Edwards, 1972) impose a mean shear rate by applying shifted periodic boundary conditions across planes normal to the shear gradient direction, $y$; particles crossing such a plane are incremented in velocity by an amount $\Delta v = \pm\dot\gamma \Lambda_y$. (They are also incremented in $x$-position by
a corresponding, time-dependent amount.) The boundary conditions across sample boundaries normal to $x$ and $z$ are the standard periodic ones. LEBC have the effect of setting up a mean shear rate $\dot\gamma$: but note that the system itself, not the boundary conditions, decide whether this shear is uniformly distributed within the sample. Specifically, LEBC in molecular dynamics does not prevent ``shear banding'' (i.e., formation of layerwise regions of different $\dot\gamma(y)$) so long as these respect the imposed {\em mean} shear rate $\dot\gamma = \Lambda_y^{-1}\int\dot\gamma(y)\,dy$.  

Wagner \& Pagonabarraga (2002) not only showed how to extend the LEBC scheme to lattice Boltzmann -- a highly nontrivial task in view of the underlying lattice -- but also showed how to include additional Lees Edwards planes. These again lie normal to $y$; if there are $n$ such planes in total, the velocity increment across each is now $\pm\dot\gamma\Lambda_y/n$. Within molecular dynamics this would be a pointless exercise in coding, but the effect in LB is to refer the physical fluid velocity within each of the $n$ slabs to a lattice that is, for a linear flow profile, moving with the velocity at the centre of that slab. This increases by a factor $n$ the maximum velocity gradient that can be modelled in LB, circumventing the scaling problem raised above. Crucially, although slabs are comoving with the {\em linear} flow profile ($\dot\gamma =$ constant) that would arise for a Newtonian fluid, the method makes no direct assumption about the {\em actual} flow profile $\dot\gamma(y)$. However, if this deviates strongly from a uniform shear rate, the fluid velocity in a certain slab could violate the requirement of $v<v_{max}$ in the lattice frame. In systems undergoing shear banding it might therefore be necessary to have an adaptive algorithm controlling the velocities of different slabs. But for the problem of sheared binary fluids of equal viscosity, shear banding is neither expected, nor observed in simulation, and the problem does not occur.

The main limitation of this approach then arises from the lattice implementation of the LEBC themselves. The velocity jump across an LE plane is small, and thus does not give an integer displacement of the lattice in each timestep. Hence there is an interpolation required to allow information from the distribution functions to stream across a given LE plane onto the mismatched lattice plane on the far side. Additionally one has to somehow increment the fluid velocity within the information that gets passed. Since the velocity set is discrete, this cannot easily be done at the level of the individual populations $f_i$ and velocities ${\bf c}_i$. Wagner \& Pagonabarraga (2002) simplify matters by applying the required shift only to the equilibrium distribution, that is, they write
\begin{equation}
f_i' = f_i + f_i^0({\bf v} + \Delta {\bf v}) - f_i^0({\bf v}) \label{wagpag}
\end{equation}
where $f_i'$ is a distribution function after passing across the plane, $f_i$ its original value, and $f_i^0$ its equilibrium value. (Note that there is no shift for velocities ${\bf c}_i$ which do not cut the plane.) A similar shift is applied to the distribution function $g_i$ that governs the binary fluid order parameter. The errors that arise from using Eq.\ref{wagpag} are, at any given timestep, demonstrably small (Wagner \& Pagonabarraga, 2002). However, they are not zero and, in combination with the interpolation effects, can build up into artefacts when integrated over long time periods. The latter problem seems to be more severe for the $g_i$ than for the $f_i$: they mainly affect the order parameter sector. Note that this sort of problem is not completely absent in systems where shear is imposed by moving solid walls at the boundaries, but clearly it is of greater concern in the case of scale-up when one wants to add additional boundary planes within the bulk of the sample.

A first-principles approach to this scale-up issue involves redesigning the implementation of the LEBC so that the streaming of information across LE planes more accurately reflects the Lees Edwards prescription at the molecular level. This work is ongoing. Meanwhile, we have tried various mitigating measures; below we describe both the artefacts found, and some of those measures.

\begin{figure}
\begin{center}
\subfigure[]{\includegraphics[%
  scale=0.3]{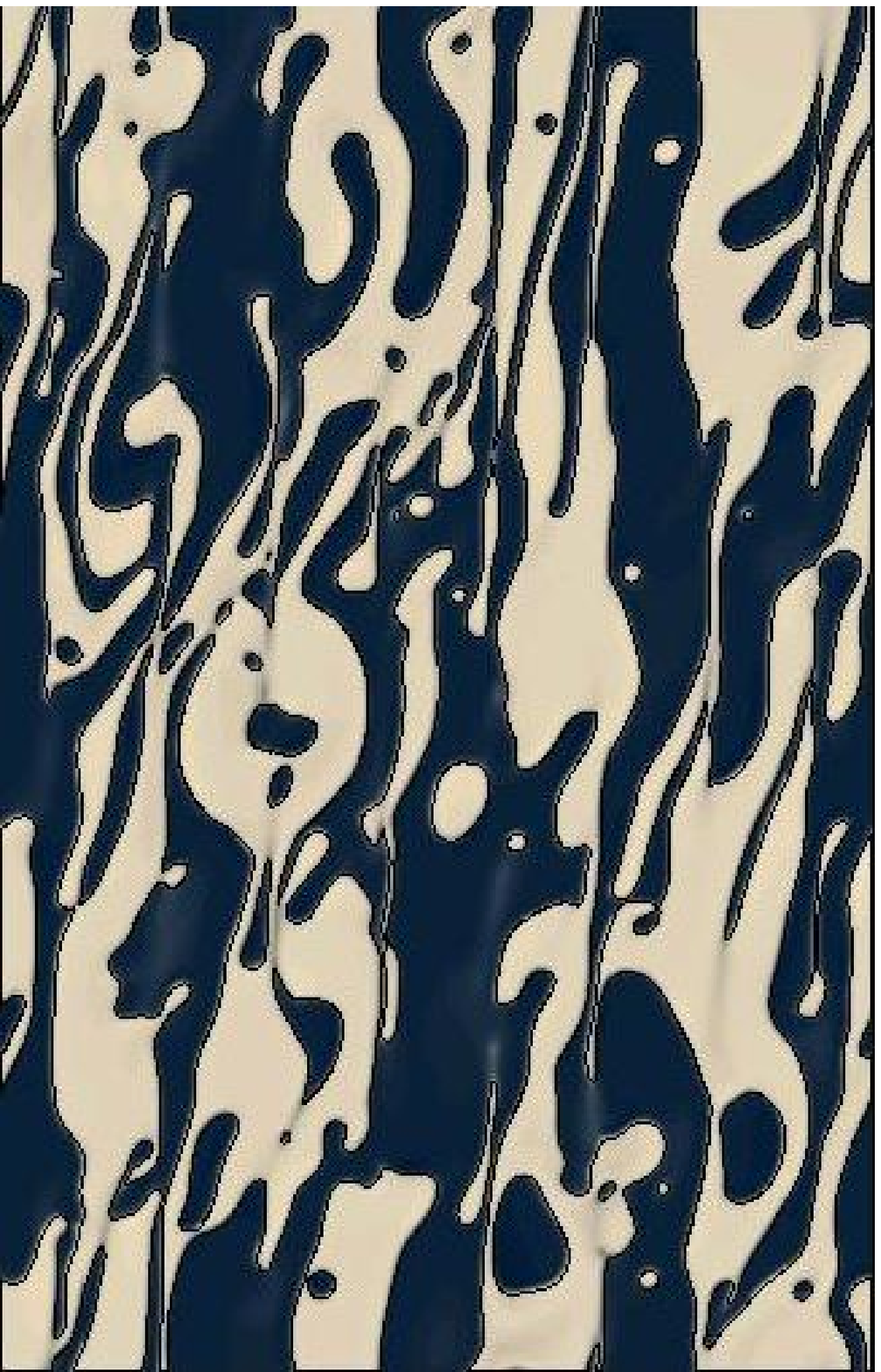}}\hspace{1cm}\subfigure[]{\includegraphics[%
  scale=0.3]{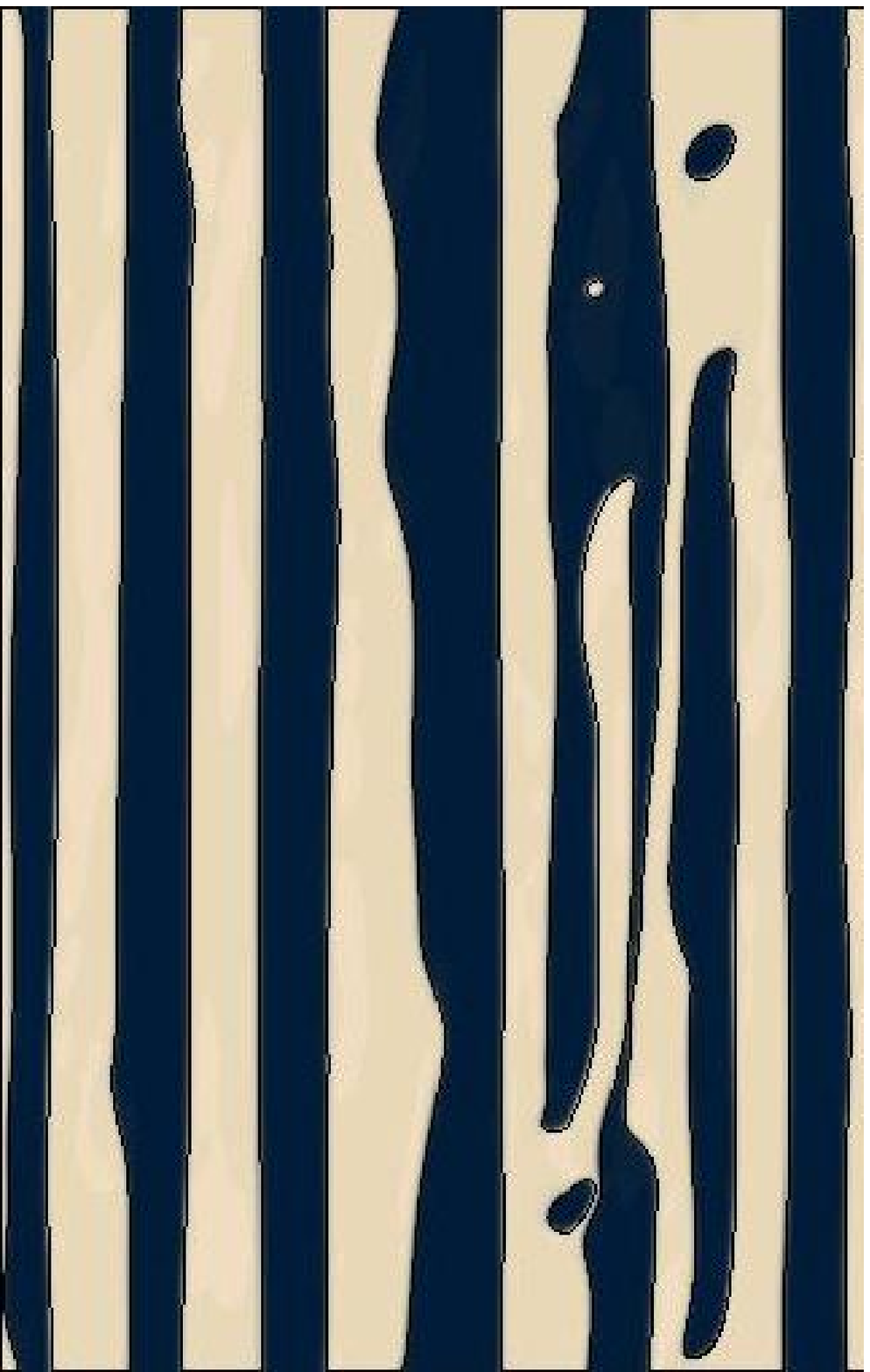}}\hspace{1cm}\subfigure[]{\includegraphics[%
  scale=0.3]{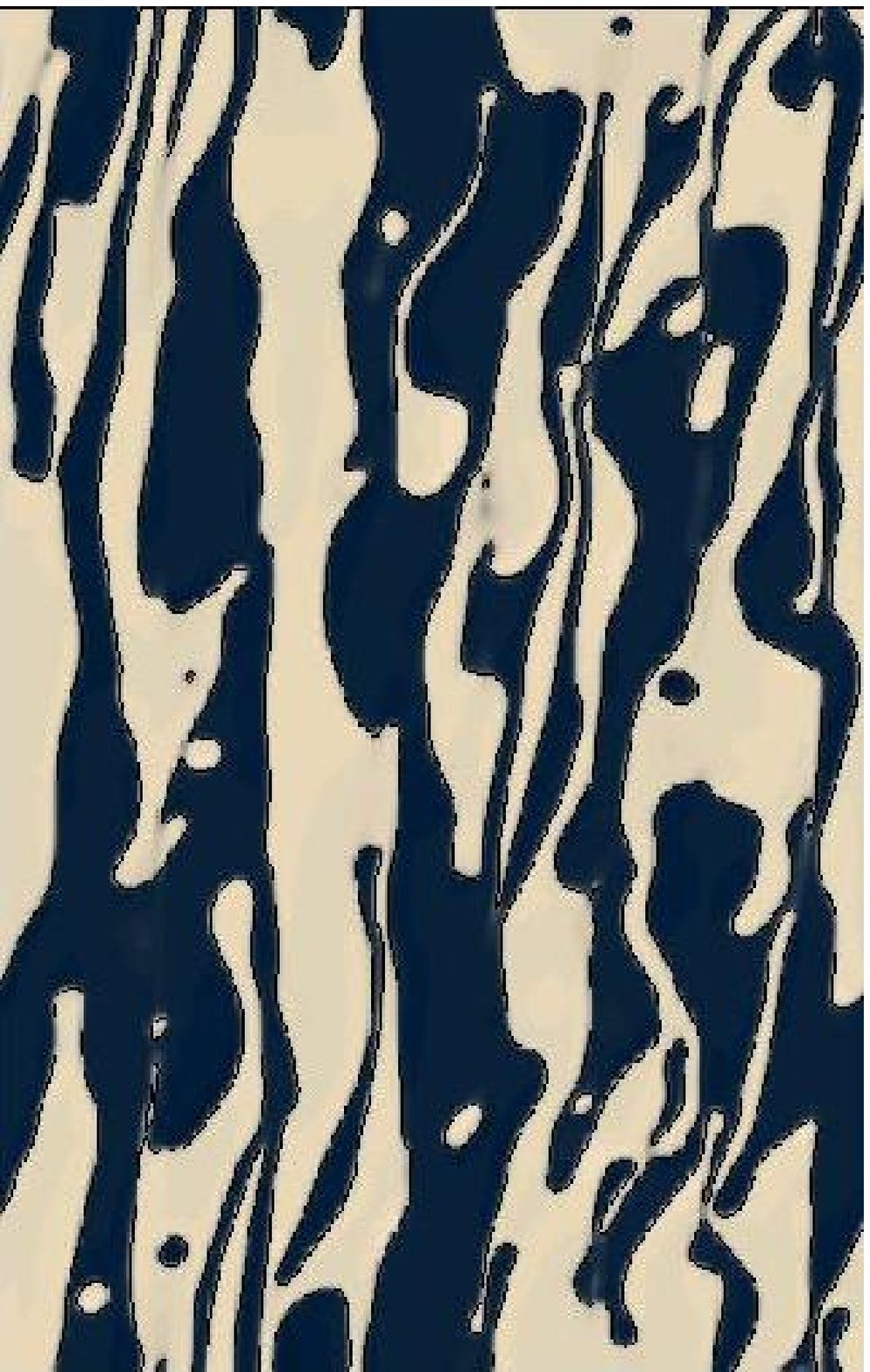}}
\end{center}

\caption{\label{fig:wogan3-phi}The positions of the LE planes are
clearly visible in the order parameter field at $t=100,000$ in (a) and
(b) but less so in (c), although some additional diffusivity on the planes remains visible. The simulation parameters were (a):
$n=8$, $\Delta v=0.1$, $\dot{\gamma}=0.0025$;
(b): $n=16$, $\Delta v=0.05$,
$\dot{\gamma}=0.0025$; (c): $n=8$,
$\Delta v =0.08$, $\dot{\gamma}=0.002$; all:
$(\Lambda_{x},\Lambda_{y})=(500,320)$,
$-A=B=0.00625$, $\kappa=0.004$, $\eta=0.025$, $M=4.0$ and $\rho=1$. }
\end{figure}

We noticed two inter-related artifacts associated with the LE planes, one seeming to restrict, and the other to promote, the
coarsening of fluid domains. The first occurred when using relatively high
values of the velocity jump $\Delta v$ across the planes (say $\Delta v>0.1$) which gave,
nevertheless, fluid velocities on either side of the plane that were
less than the commonly accepted maximum ($v< v_{max}=c_s/10 \simeq 0.058$). In such cases the locations of the LE planes were
clearly visible in the order parameter field and it appeared that the
fluid was being mixed more vigorously along these planes than
elsewhere in the simulation. This effect is illustrated in Figure
\ref{fig:wogan3-phi} (a). The growth of the domain sizes in such two dimensional
simulations reached a saturating steady state in each direction, but
this cannot be considered physically reliable due to the
visible artefact of increased fluid-fluid mixing occurring along
the LE planes. Broadly similar effects (not shown) were found in three dimensions.

The second, closely related, artefact was more evident when $\Delta v$ was reduced, so
that say $\Delta v< v_{max}/10 \simeq 0.06$. 
In these cases the growing domains had a tendency
to align themselves along the LE planes, not just locally (as already visible in  Figure
\ref{fig:wogan3-phi} (a)) but across the full extent of the simulation box.
This additional alignment was enough to promote
domain wrap-around in the direction of the shear flow. The latter is a finite size artefact (discussed below) but the interface lock-in onto LE planes greatly increases the damage done by it. 
Indeed, in some extreme cases, with an
even number of LE planes, the fluid reached a steady state in
which each domain wrapped once around the lattice in the
direction of the flow to connect with its own tail, with the interfaces aligned closely to the
positions of the LE planes. (Using an odd number of LE planes helps to
reduce, but does not eliminate, this effect.)
The resulting multi-layer steady state, such as that of Figure \ref{fig:wogan3-phi} (b), could be a valid result for a finite sized system under sheared periodic boundary conditions -- but it must be treated with great suspicion when, as here, the number and location of the interfaces closely matches those of the LE planes used.

The LB parameters in Figure \ref{fig:wogan3-phi} (a,b) are chosen to emphasise, rather than minimise, the visibility of the lock-in problem. For other parameter values the effects can be much smaller (Figure \ref{fig:wogan3-phi} (c)) and might easily be overlooked, but their cumulative effect still remains a matter of concern for quantitative work. We noticed that these artifacts became more pronounced as
simulations progressed, but always took quite some time to develop. 

A first strategy for minimizing the problem was simply to increase the number of planes while simultaneously reducing the
values of $\Delta v$ such that $\dot{\gamma}$ remains constant. This
was not satisfactory since, as foreseen by Wagner \& Pagonabarraga (2002), the numerical error
created by their scheme grows rather than vanishes in this limit. (Each plane slightly degrades the streaming of $f_i$ information even if $\Delta v = 0$.) Thus,  although the strength of the lock-in on any particular plane was
reduced by this method, we could not be confident of the overall accuracy of the simulation.

Secondly, we improved the accuracy of the interpolation of
the populations $f_{i}$ across the LE plane from linear to cubic
spline, but unfortunately this had little or no beneficial effect on
the interface lock-in. 

Thirdly, in order to reduce the time available for the artifacts to develop,
we periodically translated the positions of the LE planes normal to their own direction through a
specified number of lattice units (leaving the fluid untranslated). This was done by re-mapping the $f_i,g_i$ data, which itself introduces some interpolation errors. (But since the jumps of the planes are infrequent such additional errors are probably minor.) This plane-jumping
technique reduced the time available for alignment of fluid interfaces
with LE planes, since no such plane remains in the same place for very long. 
We found that even on very long and thin lattices, for example with
$\left(\Lambda_{x},\Lambda_{y}\right)=\left(1000,160\right)$, the
simulations still suffered from the finite-size effect of fluid domain
wrap-around. Of course, this could be the correct physics for the chosen periodic simulation box and may not be an artefact from lock-in; indeed, for lattices as large as
$\left(\Lambda_{x},\Lambda_{y}\right)=\left(2000,160\right)$ we
observed the longest length scale still to be growing, albeit slowly,
right up to our maximum simulation time of $t=300,000$. The plane-jumping technique therefore remains a possibly useful component in overcoming the scale-up problem. 

Fourthly, and in a similar vein, tests were run in which a small, constant overall velocity was imposed in the direction normal to the LE planes by initialising the fluid with a nonzero $y$ momentum. (This induces a small
steady $x$ acceleration, which could be removed by a suitable body force.) The hope was that, by drifting the entire fluid configuration sideways across the LE planes at constant velocity, there would again not be long enough for any lock-in between fluid-fluid interfaces and the LE planes to develop. Instead though, these simulations exposed a more general drawback with the LEBC scheme of Wagner \& Pagonabarraga (2002): we found that this was not Galilean invariant. Indeed, the observed velocity profile rapidly changed from the constant
shear rate expected (and seen, in the absence of transverse momentum) into something more akin to a step function. This has motivated further work on the LEBC issue, which is still ongoing. Preliminary indications are that the Galilean invariance problem can be solved within the broad framework of the Wagner--Pagonabarraga scheme, so long as some specific improvements are made to the streaming across planes of $f_i,g_i$ information. In addition, these improvements decrease (but do not wholly eliminate) the errors arising at the LE planes themselves; in combination with the drifting or the plane-hopping method, we hope that this development may soon allow production run work to begin.

\begin{figure}
\begin{center}\includegraphics[%
  scale=0.8]{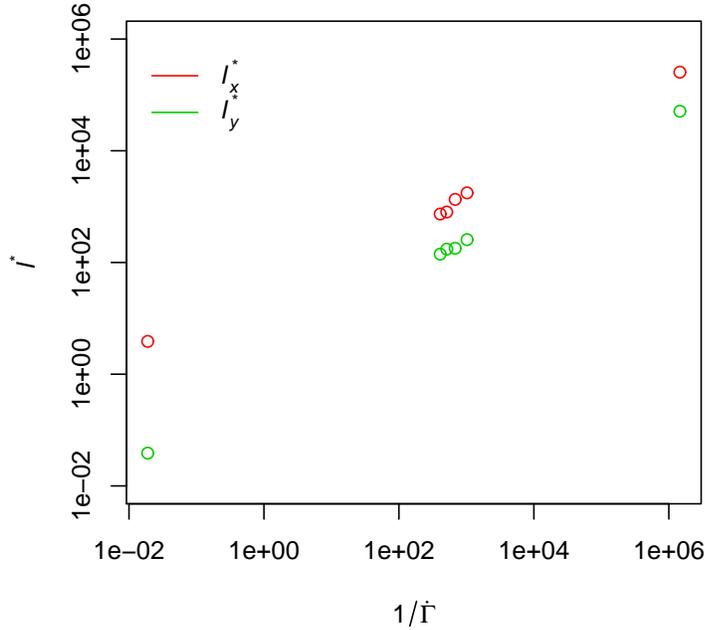}\end{center}

\caption{\label{fig:scaling} Preliminary data for near-steady-state length scales versus shear rate measured from 2D simulations. Data is nondimensionalised via characteristic parameters $L_0,T_0$ of the coarsening problem.}
\end{figure}

As a preliminary indication of what might be achieved once our scale-up problem with LE planes is surmounted, we show in Figure \ref{fig:scaling} results selected from that part of our 2D data that seemed least affected by the
above-mentioned lock-in artifacts and/or the resulting finite-size limitations. In the best 2D runs, one can discern near-steady-state length scales $L_{x}$ and $L_{y}$, defined through the order parameter field as follows:
\begin{equation}
\frac{\Lambda_x}{L_x} = \frac{1}{m}\int dA (\partial\phi/\partial x)^2 \label{lengthdef}
\end{equation}
where the area integral is over the simulation domain, and $\Lambda_y m$ is the value of the same integral for an infinitely long domain of width $\Lambda_y$ containing a single
interface perpendicular to the $x$ direction. (The expression for $L_y$ is found by interchanging $x$ and $y$; we restrict attention to the symmetric case with equal amounts of the two fluids. The generalisation to three dimensions is straightforward but messy.) With these definitions, the right hand side of Eq.\ref{lengthdef} is a measure of the number of interfaces traversed on passing through the domain along the $x$ direction, and $L_x$ accordingly gives a characteristic distance between these. (Geometrical factors involving the orientation of the interface complicate this
interpretation slightly but this need not concern us here.)

The resulting length scales are only `near' steady-state because there are large temporal fluctuations; and in many cases nucleation into a `wrapped' state, clearly limited by finite-size effects, occurs subsequent to the period of near-steady flow during which the $L$'s are defined. We have plotted in Figure \ref{fig:scaling} the resulting dimensionless lengths $l_{x,y}^{*}=L_{x,y}/L_{0}$ against the
inverse of the dimensionless shear rate
$\dot{\Gamma}=T_{0}\dot{\gamma}$, where the scaling parameters
$T_{0}$ and $L_{0}$ are those of the coarsening problem as defined previously. We make no attempt at power-law fits, but draw attention to the extremely wide range of scaled lengths and times spanned by this plot. Our data reduction method follows that of Kendon \etal (2001) although we have only two data points per run rather than a segment of the coarsening curve $l(t/T_0)$
as considered there for the unsheared case.

\subsection{Aspect ratio scaling}

As mentioned previously, there are theories that predict three growing length scales with divergent ratios for binary fluids under shear. That is, they predict (modulo some logarithmic corrections) $L_x \sim t^a, L_y\sim t^b, L_z\sim t^c$ with $a,b,c$ different exponents; see Rapapa \& Bray (1999) and Bray (2002). Whether these theories are correct, or whether coarsening instead ceases at some finite capillary or Reynolds number (Doi \& Ohta 1991, Cates \etal 1999), such predictions point to a second scale-up problem distinct from the limit on lattice-frame fluid velocities addressed above by the LEBC approach. 

\begin{figure}
\begin{center}
\includegraphics[width=12cm]{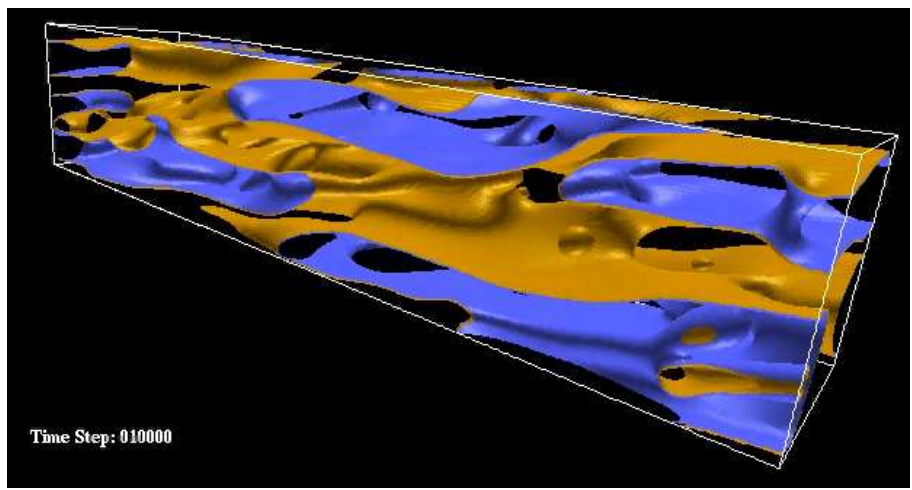}
\end{center}
\caption{\label{shearmovie} Fluid configuration late in a run showing binary fluids under shear (online movie shows the time development before and after this point). Parameter values $-A = B = 0.00625, \kappa = 0.004, \eta = 0.025, M  = 4.0$ as per Run018 of Kendon \etal (2001). ($M$ is order parameter mobility.) Both fluids are transparent; the two sides of the interface between them are painted blue and yellow. This configuration shows domains wrapped around the boundary conditions in the voriticity direction (into the page). MOVIE IN ONLINE JOURNAL (8.5MB): http://www.ph.ed.ac.uk/~mec/middle-03.avi} \end{figure}

Specifically, the shape of the simulation box required to efficiently minimise finite size effects in a system with three independently growing length scales is highly anisotropic. 
Figure \ref{shearmovie} shows a typical example of a three dimensional simulation on an $800\times 80\times 80$ lattice. (See also the work of Harting \etal (2004).) The domains wrap around the periodic boundary condition (so that a single domain connects with its own periodic image, creating in effect an infinite correlation length in that direction) first in the vorticity direction; this occurs about 1/3 of the way through the run, at around 8000 timesteps. At around 20000 timesteps wrap-around occurs along the velocity direction; this run was then terminated at 25000 timesteps.

Clearly a non-square cross section is required, but the optimal anisotropy for minimizing finite size effects depends on both shear-rate and the simulation size. Thus, doubling $\Lambda_y$ at fixed $\dot\gamma$ requires for best efficiency that $\Lambda_x$ and $\Lambda_z$ are increased by some factor other than two. Moreover, unless a theory like that of Rapapa \& Bray (1999) can be relied upon, it is impossible to know in advance what these scale-up factors should be. Since our ability to get meaningful results is critically dependent on controlling finite size effects, these problems (combined with the LEBC scale-up issue described above) have so far prevented us from gaining a clear quantitative insight into binary demixing under shear to match that of Kendon \etal (2001) for the unsheared case in three dimensions.

\begin{figure}
\begin{center}\subfigure[]{\includegraphics[%
  scale=0.5]{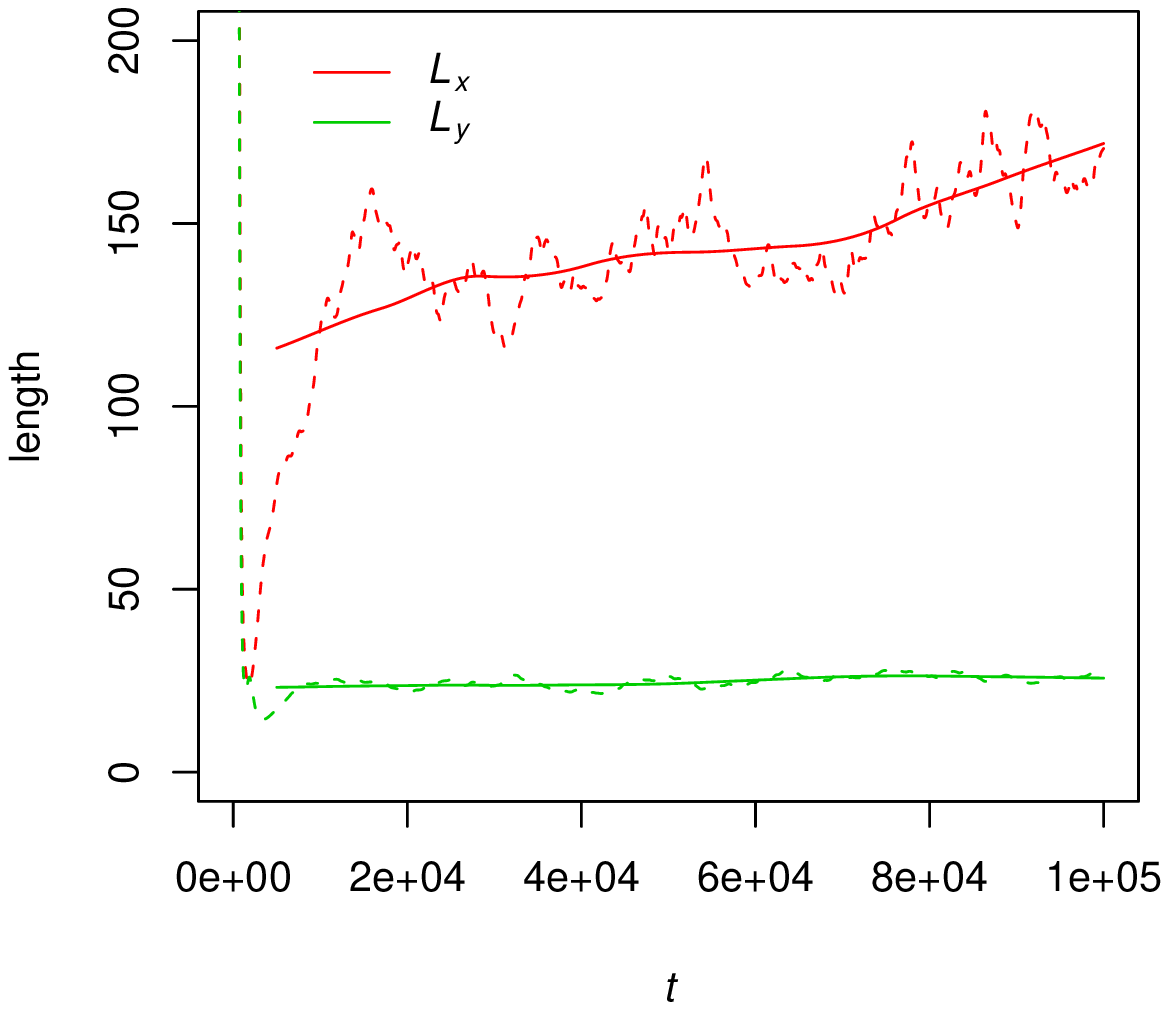}}\subfigure[]{\includegraphics[%
  scale=0.5]{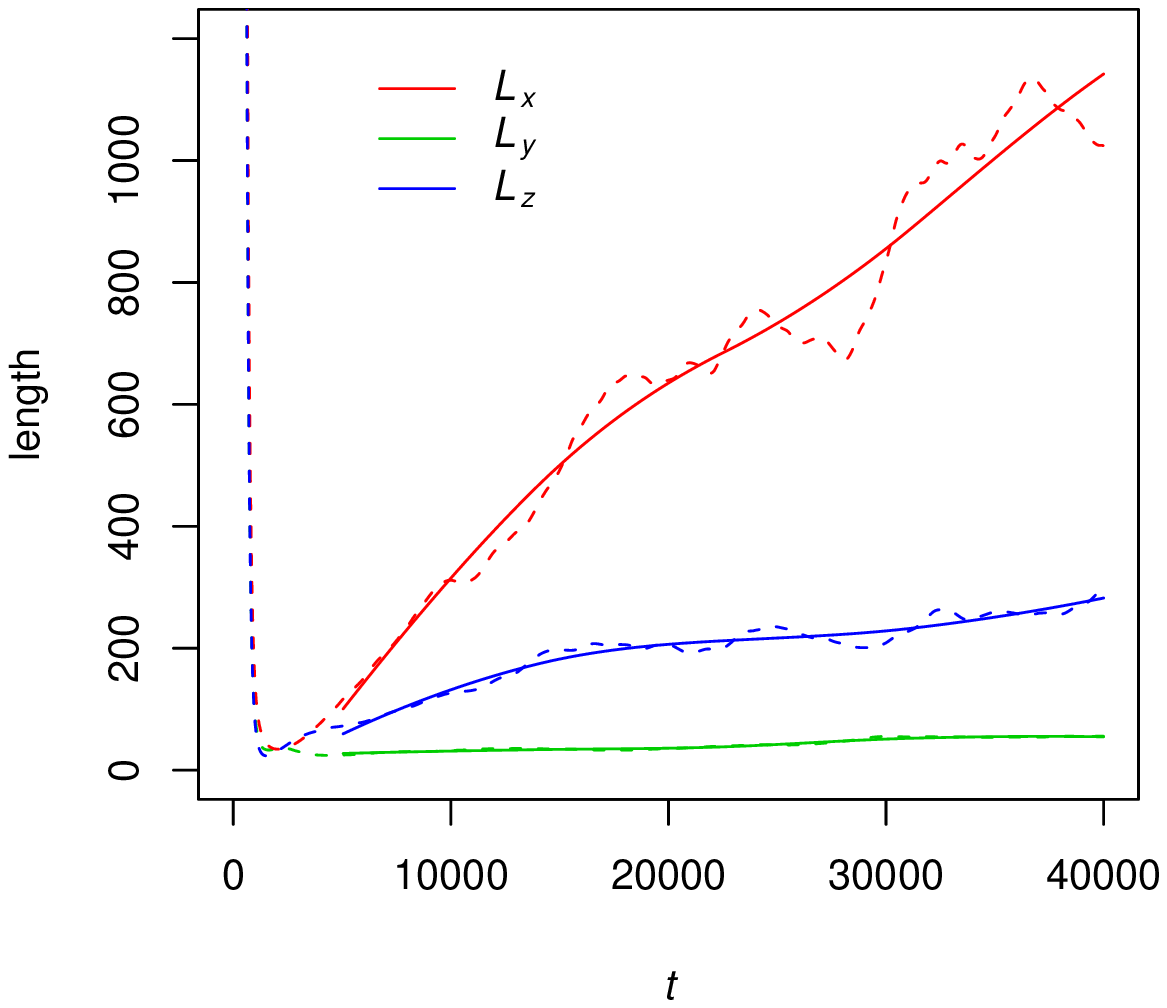}}\end{center}

\caption{\label{fig:MikesFig4} Smoothed and raw curves of
length scales. (a) $L_{x,y}(t)$ for 2D simulation of size
$(\Lambda_{x},\Lambda_{y})=(500,320)$
with $n=8$; and (b) $L_{x,y,z}(t)$ for 3D simulation of
size
$(\Lambda_{x},\Lambda_{y},\Lambda_{z})=(400,80,360)$
with $n=2$. The (common) parameters are:
$\Delta v =0.08$, $\dot{\gamma}=0.002$ $-A=B=0.00625$,
$\kappa=0.004$, $\eta=0.025$, $M=4.0$ and $\rho=1$. }
\end{figure}

Meanwhile, Figure \ref{fig:MikesFig4} shows typical $L(t)$ curves for pre-production runs in 2D and 3D. The curves are shown both as raw data and after smoothing. In the smoothed curves, a plateau can be seen at intermediate times in 2D, followed by a finite-size driven increase of the lengths as domain wrap-around occurs. (Note that, with our definitions, a wrapped configuration has $L\gg\Lambda$ in the direction of the wrap.)
Hence the identifications of near steady-state length scales, as presented in Fig. \ref{fig:scaling}, are relatively unambiguous in two dimensions.
In 3D there is a hint (only) of a similar plateau in the smoothed curve.
We expect that, if  saturation of $L$ is to be seen in three dimensions, it will be by growth of this plateau as one moves to larger system sizes. But already, the development of disparate length scales in the three directions is clear from this figure. 

\section{Mid-run alteration of sample geometry}
One tool that would make life significantly easier for such simulations would be a way of altering the size and shape of a simulation mid-run, for example if incipient wrap-around of fluid domains across (say) the vorticity direction were detected, as in Figure \ref{shearmovie}. Ultimately one would like to do this as a real-time steering operation using tools developed by the RealityGrid project; but first one has to work out how to accomplish it at all. This is possible in principle, by replicating the data of a simulation on a $\Lambda_x\times\Lambda_y\times\Lambda_z$ lattice to create one of size $\Lambda_x\times\Lambda_y\times 2\Lambda_z$. However, since the LB algorithm without noise is deterministic, nothing would be gained: the new simulation would simply calculate everything twice and deliver results with the same periodicity as before. On the other hand, the same is not true in problems where thermal or other noise (perhaps amplified by the chaotic nature of the demixing process) is dominant. In such cases the replicated system would quickly develop different behaviour in its two halves, and start to genuinely behave like a system of twice the original size.

\subsection{Example: Arrest of Rayleigh-Plateau instability by adsorbed particles}
\label{cyl}

To exemplify the principle of mid-run alteration of sample geometry, we apply the idea here to a specific problem. The question is as follows: if a cylindrical interface between two fluids is coated with a random arrangement of colloidal particles with neutral wetting characteristics (see Cates \etal 2004 and Figure \ref{bench}), does this arrest the interfacial dynamics? 

\begin{figure}
\begin{center}
\includegraphics[width=12cm]{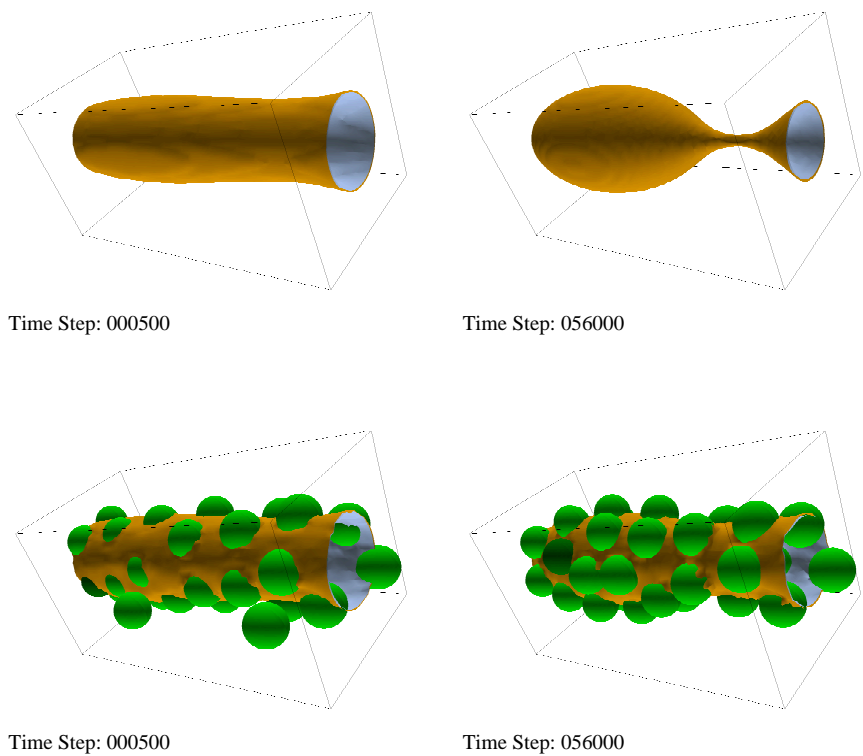}
\end{center}
\caption{\label{instability} Upper frames: evolution of a fluid cylinder undergoing Rayleigh-Plateau instability; initial state, and 56,000 timesteps after initiation. Lower frames: the same system with neutrally wetting particles coating the cylinder surface; initial state, and 56,000 timesteps after initiation. The particle-laden initial state was created by by replicating a shorter cylinder as described in the text.  LB parameters are $-A = B = 0.002, \kappa = 0.0014,\eta = 0.1, M = 2$. The colloid radius is 3.7, and $k_BT = 3.3\times 10^{-4}$. (All quantities are expressed in lattice units.) The
particle-laden simulation contains additional particles, exterior to the cylinder and not contacting its surface, which are removed from the images for clarity.} \end{figure}

A good test of this is to see whether the Rayleigh-Plateau instability, whereby a long cylinder of fluid breaks into droplets under the influence of interfacial tension, is suppressed by the presence of the adsorbed particles (Figure \ref{instability}). A technical problem then lies in creating a realistic initial state. It would be possible to simply place the particles on the interface by hand, but there is a danger of creating a state that is not the outcome of any physically realisable process (for example, it might have an unrealistically high particle density). Thus it seems preferable to mimic the dynamics of Figure \ref{bench}, and allow the colloidal particles to coat the surface under the influence of a body force acting on them. A reasonable choice, for a cylinder, is a body force that is radially directed towards the cylinder surface and linearly proportional to the distance from this surface. 

However, to allow this to happen, the cylinder of fluid has first to be stabilised {\em against} the Rayleigh-Plateau effect for long enough that the particles can arrive at its surface and coat this uniformly, before departures arise from a cylindrical shape. This can be done by mid-run intervention as follows. It is known that the instability has a minimum wavelength $\Lambda_c = 2 R$ and a most unstable wavelength $\Lambda_m = 9.02 R$, where $R$ is the cylinder radius (Faber, 1995). Therefore, by imposing periodic boundary conditions with period $\Lambda < \Lambda_c$ along the cylinder axis, the instability is switched off. (In practice, $\Lambda < 2 \Lambda_c$ is sufficient since growth rates are very slow away from $\Lambda_m$.) In such a simulation, it is then a simple matter to deliver a coating of particles to the interface in the manner just outlined. This coated cylinder can then be replicated as described above, by gluing two (or more) identical copies side by side so that the length of the simulation box is now much greater than $\Lambda_c$. As mentioned previously, noiseless evolution of this state would retain periodicity; but the instability, if present, ensures that any small-amplitude extrinsic perturbations with wavelength above $\Lambda_c$ gets amplified. Our simulations are done with thermal noise, so in principle one could wait for the required perturbation to arise spontaneously but in practice this is slow. (This fact suggests that we might get away with using a long cylinder initially rather than our replication method; but the perturbation caused by arriving particles is significantly larger than that of thermal noise so that this is not automatic.) Instead we can add a one-off perturbation to the replicated system by hand, and watch the result. We carried this out for an an initial lattice of $32^3$ with cylinder radius $R = 8$ in lattice units; the initial particle volume fraction was 20\%. About two thirds of these particles reached the surface of the cylinder within 50,000 timesteps. The system was then replicated (doubled in length) and initialized with a small amplitude perturbation (ca. 1 lattice unit in the radius of the cylinder) at wavelength $64 = 0.87 \Lambda_c$. This was then run on for 60,000 timesteps, during which there was little growth of the instability. Initialising to the same state without particles led instead to pinchoff after 56,000 timesteps. Figure \ref{instability} shows equal-time snapshots for the system with and without particles.

\section{Conclusions}
In this paper we have considered various physical and computational scaling issues that arise in the lattice Boltzmann simulation of binary fluid mixtures, and illustrated these with examples from our recent work on sheared binary fluids and colloidal particles in binary solvents. To make good use of mesoscale simulation methods one needs to appreciate the hierarchy of length- and timescales that are present in the physical system of interest, and try to achieve the same hierarchy within the simulation. It is, however, neither necessary nor generally practical to fully resolve all the time- and length scale separations present in this hierarchy, which in the real world can span many decades. Instead, one should pay attention to the major physical scales and at least ensure that these occur in the right order in the simulation. 

Computational scale-up of LB work is hampered by the various special measures that are needed to reconcile large-scale shear (which requires local Galilean invariance in the fluid frame) with the existence of an underlying computational lattice. While progress has been made, this issue is not yet fully understood and we hope to report further on it in future work. For systems under shear, strongly anisotropic simulation domains are also needed; in some cases it would be very helpful if the size and aspect ratio of these could be changed interactively. It is not clear yet whether mid-run intervention to create larger simulation boxes by replicating the data will ever be useful in that context; this seems more promising for noise-dominated and/or chaotic systems than any exhibiting simpler dynamics. Such interventions can however already be useful, for example in the creation of a complicated initial condition that would otherwise lie on the wrong side of an instability boundary. 

\begin{acknowledgements} This work was funded in part by EPSRC Grants GR/R67699 (RealityGrid) and GR/S10377. We thank Patrick Warren for useful discussions.
\end{acknowledgements}

\label{lastpage}
\end{document}